\documentclass[runningheads,envcountsect,envcountsame]{llncs}
\usepackage[utf8]{inputenc}
\usepackage[T1]{fontenc}

\usepackage{bm}\usepackage{amsmath}\usepackage{amssymb}

\usepackage{mathrsfs}\usepackage{stmaryrd}\usepackage{cleveref}\usepackage{mathtools}

\usepackage{tikz}
\usetikzlibrary{arrows.meta}    

\spnewtheorem{algorithm}[theorem]{Algorithm}{\bfseries}{\rmfamily}

\newcommand{\Cuts}{\textit{Cuts}}
\newcommand{\dom}{\textit{dom}}

\newcommand{\Bcal}{\mathcal{B}}

\newcommand{\ta}{\mathtt{a}}
\newcommand{\tb}{\mathtt{b}}
\newcommand{\tc}{\mathtt{c}}

\newcommand{\tS}{\mathtt{s}}
\newcommand{\RHS}{\rho}

\newcommand{\eqdef}{\stackrel{\text{def}}{=}}

\renewcommand{\exp}{\operatorname{\textit{exp}}}

\newcommand{\Nat}{\mathbb{N}}
\newcommand{\Rrm}{{\mathrm{R}}}

\newcommand{\subword}{\preccurlyeq}

\renewcommand{\epsilon}{\varepsilon}\renewcommand{\emptyset}{\varnothing}\renewcommand{\setminus}{\smallsetminus}

\newcommand{\pair}[1]{\langle #1 \rangle}\newcommand{\bpair}[1]{\bigl\langle #1 \bigr\rangle}

\title{On arch factorization and subword universality \\
for words and compressed words}
\titlerunning{On arch factorization and subword universality}
\author{Ph.\ Schnoebelen \and J.\ Veron}
\institute{LMF, CNRS \& ENS Paris-Saclay, France
\thanks{Work partially supported by Labex DigiCosme (project ANR-11-LABEX-0045-DIGICOSME) operated by ANR as part of the program «~Investissement d'Avenir~» Idex Paris-Saclay (ANR-11-IDEX-0003-02).}}

\begin{document}

\maketitle

\begin{abstract}
Using arch-jumping functions and properties of the arch factorization of words, we propose a new algorithm for computing the subword circular universality index of words.  We also introduce the subword universality signature for words, that leads to simple algorithms for the universality indexes of SLP-compressed words.
\end{abstract}

\section{Introduction}
\label{sec-intro}

A \emph{subword} of a given word is obtained by removing some letters
at arbitrary places. For example, $\mathtt{abba}$ is a subword of
$\mathtt{\underline{ab}racada\underline{b}r\underline{a}}$, as
witnessed by the underlined letters.  Subwords are a fundamental
notion in formal language theory and in algorithmics but they are not
as well-behaved as \emph{factors}, a special case of subwords where
the kept letters correspond to an interval inside the original
word.\footnote{Some papers use the terminology ``subwords'' for
factors, and ``scattered subwords'' or ``scattered factors'' for
subwords.  We follow~\cite{sakarovitch83}.}

Words and languages can be characterised or compared via their
subwords. For example, we can distinguish $u_1=\mathtt{nationalists}$
from $u_2=\:$\texttt{anti\-na\-tiona\-lists} by the subword
$x=\mathtt{ino}$.  Indeed, only $u_2$ has $x$ as a subword. We say
that $x$ is a \emph{distinguisher} (also, a \emph{separator}) between
$u_1$ and $u_2$.  Observe
that $\mathtt{ino}$ is a \emph{shortest} distinguisher
 between the two words.\footnote{This is a very rare situation
with the English lexicon, where different words  almost always admit
a  length-2 distinguisher. To begin with, two words can
already admit a  length-1 distinguisher unless they use exactly
the same set of letters.} In applications one may want to distinguish
between two similar DNA strings, or two traces of some program
execution: in these situations where inputs can be huge, finding a
short distinguishing subword requires efficient algorithms~\cite{simon2003}. When
considering the usual first-order logic of words (i.e., labelled linear orders), a distinguisher $x$
can be seen as a $\Sigma_1$ formula separating the two words.

\paragraph{Definability by subwords.}
These considerations led Imre Simon to the introduction of \emph{piecewise-testable}
languages in his 1972 Phd thesis~\cite{simon72,simon75}: these languages can
be defined entirely in terms of forbidden and required subwords. In
logical terms, this corresponds to $\Bcal\Sigma_1$-definability, see~\cite{DGK-ijfcs08}.
Piecewise testability is an important and fundamental concept, and it
has been extended to, among others,
trees~\cite{bojanczyk2012b,goubault2016}, picture
languages~\cite{matz98}, or words over arbitrary scattered linear
orderings~\cite{carton2018b}.

From a descriptive complexity point of view, a relevant measure is the
\emph{length of subwords} used in defining piecewise-testable
languages, or in distinguishing between two individual words.
Equivalently, the required length for these subwords is the required
number of variables for the $\Bcal\Sigma_1$ formula.
 This
measure was investigated in~\cite{KS-lmcs2019} where it is an
important new tool for bounding the complexity of decidable logic
fragments.

\paragraph{Subword universality.}
Barker, Day \textit{et al.\ } introduced the notion of subword
universality:  a word $u$ is \emph{$k$-universal} if all
words of length at most $k$ are subwords of
$u$~\cite{barker2020,day2021}. They further define the \emph{subword
universality index} $\iota(u)$ as the largest $k$ such that $u$ is
$k$-universal. Their motivations come, among others, from works in
reconstructing words from subwords~\cite{day2019} or computing edit
distance~\cite{day2021}, see also the survey in~\cite{kosche2022b}.
In~\cite{barker2020}, the authors prove several properties of
$\iota(u)$, e.g., when $u$ is a palindrome, and further introduce the
\emph{circular} subword universality index $\zeta(u)$, which is
defined as the largest $\iota(u')$ for $u'$ a conjugate of $u$.
Alternatively, $\zeta(u)$ can be seen as the subword universality
index $\iota([u]_\sim)$ for a \emph{circular word} (also called
necklace, or cyclic word), i.e., an equivalence class of words modulo
conjugacy.

While it is easy to compute $\iota(u)$, computing $\zeta(u)$ is
trickier but \cite{barker2020} proves several bounds relating
$\zeta(u)$ to the values of $\iota(u^n)$ for $n\in\Nat$. This is
leveraged in~\cite{fleischmann2021} where an $O(|u|\cdot|A|)$
algorithm computing $\zeta(u)$ is given. That algorithm is quite
indirect, with a delicate and nontrivial correctness proof.  Further
related works are~\cite{kosche2021} where, given that $\iota(u)=k$,
one is interested in all the words of length $k+1$ that
do not occur as subwords of $u$, \cite{fleischmann2022b} where one
considers words that are just a few subwords away from
$k$-universality, and~\cite{kosche2022} where the question whether $u$
has a $k$-universal factor of given length is shown to be
\textsf{NP}-complete.

\paragraph{Our contribution.}
In this paper we introduce new tools for studying subword (circular)
universality. First we focus on the arch factorizations (introduced by
H\'ebrard~\cite{hebrard91}) and show how \emph{arch jumping} functions
lead to simple proofs of combinatorial results on subword universality
indexes, allowing a new and elegant algorithm for computing
$\zeta(u)$.  These arch-jumping functions are implicit in some
published constructions and proofs (e.g.,
in~\cite{fleischer2018,fleischmann2021,kosche2021})
but studying
them explicitly brings simplifications and improved clarity.

In a second part we give bilinear-time algorithms that compute the
universality indexes $\iota$ and $\zeta$ for compressed words. This is
done by introducing a compact \emph{subword universality signature}
that can be computed compositionally.  These algorithms and the
underlying ideas can be useful in the situations we mentioned earlier
since long DNA strings or program execution traces are usually very
repetitive, so that handling them in compressed form can entail huge
savings in both memory and communication time.

More generally this is part of a research program on algorithms and
logics for computing and reasoning about
subwords~\cite{KS-msttocs,HSZ-lics2017,KS-lmcs2019,ghkks-ideals}. In
that area, handling
words in compressed form raises additional difficulties. For example it is
not known whether one can compute efficiently the length of a shortest
distinguisher between two
compressed words. Let us recall here
that reasoning on subwords is usually harder than reasoning on factors, and
this is indeed true for compressed words:
While deciding whether a compressed $X$ is a \emph{factor} of a
compressed $Y$ is polynomial-time, deciding whether $X$ is a
\emph{subword} of $Y$ is intractable (in \textsf{PSPACE} and
\textsf{PP}-hard, see~\cite[Sect.~8]{lohrey2012}). However, in the
special case where one among $X$ or $Y$ is a \emph{power word}, i.e.,
a compressed word with restricted nesting of concatenation and
exponentiation, the subword relation is polynomial-time, a result
crucial for the algorithms in~\cite{phs-csl2021} where one handles
exponentially long program executions in compressed forms.

\paragraph{Outline of the paper.}
\Cref{sec-basics} recalls all the necessary definitions for subwords and
universality indexes.  \Cref{sec-alpha} introduces the arch-jumping
functions, relates them to universality indexes and proves some basic
combinatorial results.  Then \Cref{sec-zeta-algo} provides a simple
algorithm for the circular universality index.  In
\Cref{sec-signatures} we introduce the subword universality signature of
words and show how they can be computed compositionally.
Finally \Cref{sec-SLPs} considers SLP-compressed words and
their subword universality indexes.

 \section{Basic notions}
\label{sec-basics}

\paragraph{Words and subwords.}
Let $A = \{\ta,\tb,\ldots\}$ be a finite alphabet.  We write
$u$, $v$, $w$, $s$, $t$, $x$, $y\ldots$ for words in $A^*$. Concatenation is denoted
multiplicatively while $\epsilon$ denotes the empty word.  When $u = u_1 u_2 u_3$
we say that $u_1$ is a \emph{prefix}, $u_2$ is a \emph{factor}, and $u_3$ is a
\emph{suffix}, of $u$.  When $u = v w$ we may write $v^{-1}u$ to denote $w$,
the suffix of $u$ one obtains after removing its $v$ prefix. When $u = v_0 w_1
v_1 w_2 \cdots w_n v_n$, the concatenation $w_1w_2\cdots w_n$ is a
\emph{subword} of $u$, i.e., a subsequence obtained from $u$ by removing some
of its letters (possibly none, possibly all). We write $u\subword v$ when $u$
is a subword of $v$.

A word $u = a_1\cdots a_\ell$ has length $\ell$, written $|u| = \ell$, and we let
$A(u)\eqdef  \{a_1,\ldots,a_\ell\}$  denote its alphabet, a subset of $A$. We
let $\Cuts(u) = \{0,1,\ldots,\ell\}\subseteq\Nat$ denote the set of
\emph{cutting positions inside $u$}, i.e., positions between $u$'s letters,
 where $u$
can be split: for $0\leq i\leq j\leq \ell$, we let $u(i,j)$ denote the factor
$a_{i+1}a_{i+2}\cdots a_{j}$.  With this notation, $u(0,j)$ is $u$'s prefix of
length $j$, and $u(i,\ell)$ is the suffix $(u(0,i))^{-1}u$. Note also that
$u(i,i) = \epsilon$ and $u(i,j) = u(i,k) \, u(k,j)$ whenever the factors are
defined.  If $u = u_1u_2$, we say that $u_2u_1$ is a \emph{conjugate} of
$u$. For $i\in\Cuts(u)$, the $i$-th conjugate of $u$ is $u(i,\ell) \, u(0,i)$ and
is denoted by $u^{\sim i}$. Finally $u^\Rrm \eqdef a_\ell \cdots a_1$ denotes
the \emph{mirror} of $u$.

\paragraph{Rich words and arch factorizations.}
A  word $u\in A^*$ is \emph{rich} if it contains at least one
occurrence of each letter $a\in A$, otherwise we say that it is
\emph{incomplete}.  A rich word having no rich strict prefix is an
\emph{arch}.
The mirror of an arch is called a \emph{co-arch} (it is generally not an arch). Observe that an arch (or a co-arch)
necessarily ends (respectively, starts) with a letter that occurs
only once in it.

The \emph{arch factorization} of $u$, introduced by Hebrard~\cite{hebrard91},
is a decomposition $u = s_1\cdots s_m\cdot r$ of $u$ into $m+1$ factors given
by the following:\\
--- if $u$ is not rich then $m = 0$ and $r = u$,
\\
--- otherwise let $s_1$ be the shortest prefix of $u$ that is rich (it is
an arch) and let
$s_2,\ldots,s_m, r$ be the arch factorization of the suffix $(s_1)^{-1}u$.
\\

We write $r(u)$ for the last factor in $u$'s factorization,  called the \emph{rest} of $u$.  For example, with $A =
\{\ta,\tb,\tc\}$, the arch factorization of $u_{\text{ex}} = \mathtt{b a c c a b b c b a a b a c b a}$ is $\mathtt{b a
c}\cdot\mathtt{c a b}\cdot\mathtt{b c b a}\cdot\mathtt{a b a c}\cdot\mathtt{b a}$, with $m = 4$ and $r(u_{\text{ex}}) = \tb\ta$.
Thus the arch factorization is a leftmost decomposition of $u$ into arches, with a final rest $r(u)$.

There is a symmetric notion of co-arch factorization where one factors $u$ as $u=r'\cdot s'_1\cdots s'_m$ such that $r'$ is
incomplete and every $s'_i$ is a co-arch, i.e., a rich factor whose first letter occurs only once.

All the above notions assume a given underlying alphabet $A$, and we should speak more precisely of ``$A$-rich'' words,
``$A$-arches'', or ``rest $r_A(u)$''. When $A$ is understood, we retain the simpler terminology and notation.

\paragraph{Subword universality.}
In~\cite{barker2020}, Barker et al.\ define the \emph{subword
universality index} of a word $u$, denoted $\iota_A(u)$, or just
$\iota(u)$, as the largest $m\in\Nat$ such that any word of length
$m$ in $A^*$ is a subword of $u$.

It is clear that $\iota(u) = m$ iff the arch factorization of $u$ has $m$
arches. Hence one can compute $\iota(u)$ in linear time
simply by scanning $u$ from left to right, keeping track of letter appearances
in consecutive arches, and counting the arches~\cite[Prop.~10]{barker2020}.
Using that scanning algorithm for $\iota$, one sees that the
following equalities hold for all words $u,v$:
\begin{xalignat}{2}
\label{eq-iota-concat}
\iota(u \, v) &= \iota(u) + \iota\bigl( r(u) v \bigr) \:,
&
r(u \, v) &= r\bigl( r(u) v \bigr) \:.
\end{xalignat}

Barker et al.\ further define the \emph{circular subword universality index} of
$u$, denoted $\zeta(u)$, as the largest $\iota(u')$ for $u'$ a conjugate of
$u$. Obviously, one always has $\zeta(u)\geq \iota(u)$. Note that $\zeta(u)$
can be strictly larger that $\iota(u)$, e.g., with $A = \{\ta,\tb\}$ and $u =
\mathtt{a a b b}$ one has $\iota(u) = 1$ and $\zeta(u) = 2$. These descriptive
complexity measures are invariant under mirroring of words, i.e.,
$\iota(u^\Rrm)=\iota(u)$ and $\zeta(u^\Rrm)=\zeta(u)$, and
monotonic w.r.t.\ the subword ordering:
\begin{equation}
\label{eq-iota-zeta-mono}
    u\subword v \implies \iota(u)\leq\iota(v) \land \zeta(u)\leq\zeta(v)\:.
\end{equation}

The behaviour of $\zeta$ can be deceptive. For example, while
$\iota$ is superadditive, i.e., $\iota(u v)\geq \iota(u)+\iota(v)$
---just combine \cref{eq-iota-concat,eq-iota-zeta-mono}--- we observe
that $\zeta(u v)<\zeta(u)+\zeta(v)$ can happen, e.g., with $u =
\ta\tb$ and $v = \tb\tb\ta\ta$.

\section{Arch-jumping functions and universality indexes}
\label{sec-alpha}

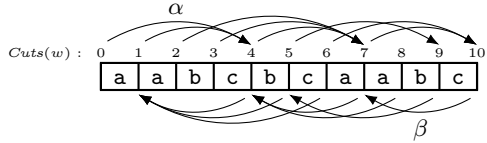
\begin{figure}[htbp]
\begin{center}
\begin{tikzpicture}[auto,x=5mm,y=3mm,anchor=mid,baseline,node distance=1.3em]
\def\sqwidth{1}\def\sqbot{-0.5}\def\sqtop{0.7}\def\posh{1.1}\def\alphh{1.5}\def\alphph{-0.7}

{\tikzstyle{every node}=[]
\node at (0.5*\sqwidth,0) {$\ta$};
\node at (1.5*\sqwidth,0) {$\ta$};
\node at (2.5*\sqwidth,0) {$\tb$};
\node at (3.5*\sqwidth,0) {$\tc$};
\node at (4.5*\sqwidth,0) {$\tb$};
\node at (5.5*\sqwidth,0) {$\tc$};
\node at (6.5*\sqwidth,0) {$\ta$};
\node at (7.5*\sqwidth,0) {$\ta$};
\node at (8.5*\sqwidth,0) {$\tb$};
\node at (9.5*\sqwidth,0) {$\tc$};
}
{\tikzstyle{every path}=[draw,thick,-]
\path (0,\sqbot) -- (10*\sqwidth,\sqbot) -- (10*\sqwidth,\sqtop) -- (0,\sqtop) -- cycle;
\foreach \i in {0,...,10} \path (\i,\sqbot) -- (\i,\sqtop);
}
{\tikzstyle{every node}=[node font=\tiny]
\node at (0*\sqwidth,\posh) {$0$};
\node at (1*\sqwidth,\posh) {$1$};
\node at (2*\sqwidth,\posh) {$2$};
\node at (3*\sqwidth,\posh) {$3$};
\node at (4*\sqwidth,\posh) {$4$};
\node at (5*\sqwidth,\posh) {$5$};
\node at (6*\sqwidth,\posh) {$6$};
\node at (7*\sqwidth,\posh) {$7$};
\node at (8*\sqwidth,\posh) {$8$};
\node at (9*\sqwidth,\posh) {$9$};
\node at (10*\sqwidth,\posh) {$10$};

\node at (-1.5*\sqwidth,\posh) {$\Cuts(w):$};
}

{\tikzstyle{every path}=[-Latex,shorten <=1mm]
\path (0*\sqwidth,\alphh) edge [thin,bend left=30] node [above] {$\alpha$} (4*\sqwidth,\alphh);
\path (1*\sqwidth,\alphh) edge [bend left=30] (4*\sqwidth,\alphh);
\path (2*\sqwidth,\alphh) edge [bend left=30] (7*\sqwidth,\alphh);
\path (3*\sqwidth,\alphh) edge [bend left=30] (7*\sqwidth,\alphh);
\path (4*\sqwidth,\alphh) edge [bend left=30] (7*\sqwidth,\alphh);
\path (5*\sqwidth,\alphh) edge [bend left=30] (9*\sqwidth,\alphh);
\path (6*\sqwidth,\alphh) edge [bend left=30] (10*\sqwidth,\alphh);
\path (7*\sqwidth,\alphh) edge [bend left=30] (10*\sqwidth,\alphh);

\path (10*\sqwidth,\alphph) edge [bend left=30] node [below] {$\beta$} (7*\sqwidth,\alphph);
\path (9*\sqwidth,\alphph) edge [bend left=30] (5*\sqwidth,\alphph);
\path (8*\sqwidth,\alphph) edge [bend left=30] (4*\sqwidth,\alphph);
\path (7*\sqwidth,\alphph) edge [bend left=30] (4*\sqwidth,\alphph);
\path (6*\sqwidth,\alphph) edge [bend left=30] (1*\sqwidth,\alphph);
\path (5*\sqwidth,\alphph) edge [bend left=30] (1*\sqwidth,\alphph);
\path (4*\sqwidth,\alphph) edge [bend left=30] (1*\sqwidth,\alphph);
}

\end{tikzpicture}
 \end{center}
\caption{Arch-jumping functions $\alpha,\beta$ for $A=\{\ta,\tb,\tc\}$ and $w=\mathtt{a a b c b c a a b c}$.}
\label{fig-alpha-maps}
\end{figure}

Let us fix a word $w = a_1 a_2\cdots a_L$ of length $L$.  We now introduce the
$\alpha$ and $\beta$ \emph{arch-jumping functions} that describe the reading of an
arch starting from some position inside $w$.  For $i\in\Cuts(w)$, we let
\begin{xalignat*}{2}
\alpha(i)  & = \min \{j~|~A\bigl(w(i,j)\bigr)=A\},
&
\beta(j) & = \max \{i~|~A\bigl(w(i,j)\bigr)=A\}.
\end{xalignat*}
These are partial functions: $\alpha(i)$ and $\beta(j)$ are undefined when
$w(i,L)$ or, respectively, $w(0,j)$, does not contain all the letters from
$A$. See \Cref{fig-alpha-maps} for an illustration.

The following properties are easily seen to hold for all $i,j\in\dom(\alpha)$:
\begin{xalignat}{2}
 \alpha(i) &\geq i+|A| \:,
&
 i\leq j \implies & \alpha(i)\leq \alpha(j) \:,
\\
 \beta(\alpha (i)) &\geq i     \:,
&
 \alpha(\beta(\alpha(i))) &= \alpha(i) \:.
\end{xalignat}
Since $\beta$ is a mirror version of $\alpha$, it enjoys similar properties
that we won't spell out here.

\begin{remark}
As will be seen in the rest of this section, the arch jumping
functions are a natural and convenient tool for reasoning about arch
factorizations. Similar concepts can certainly be found in the
literature. Already in~\cite{hebrard91}, H\'ebrard writes $p(n)$ for
what we write $\alpha^n(0)$, i.e., the $n$-times iteration
$\alpha(\alpha(\cdots(\alpha(0))\cdots))$ of $\alpha$ on $0$: the
starting point for the $p(n)$'s is
fixed, not variable. In~\cite{fleischer2018}, Fleischer and Kufleitner
use rankers like $\mathsf{X}_a$ and $\mathsf{Y}_b$ to jump from a
current position in a word to the next (or previous) occurrence of a
given letter, here $a$ and $b$: this can specialise to our $\alpha$
and $\beta$ if one knows what is the last letter of the upcoming
arch. In~\cite{kosche2021} $\texttt{minArch}$ corresponds exactly to
our $\alpha$, but there $\texttt{minArch}$ is a data
structure used to store information, not a notational tool for
reasoning algebraically about arches.
\end{remark}

\subsection{Subword universality index via jumping functions}

The connection between the jumping function $\alpha$ and the subword
universality index $\iota(w)$ is clear:
\begin{equation}
\label{eq-iota-via-alpha}
\iota(w) = \max \bigl\{ n ~\big|~ \alpha^n(0) \text{ is defined}\bigr\}\:.
\end{equation}
For example, $w$
in \Cref{fig-alpha-maps} has $\alpha^3(0)=10=|w|$ so $\iota(w)=3$.

\begin{figure}[htbp]
\begin{center}
\begin{tikzpicture}[auto,x=5mm,y=3mm,anchor=mid,baseline,node distance=1.3em]

\def\xLz{0}
\def\xLu{10.5}
\def\xLuv{19.5}

\def\xuo{2.3};
\def\xui{4.0};
\def\xuii{6.1};
\def\xumi{\xLu-1.5};

\def\xvmo{\xLuv-1.3}
\def\xvmi{\xLuv-3.1}
\def\xvmii{\xLuv-4.9}
\def\xvi{\xLu+1.8}

{\tikzstyle{every node}=[node font=\Large]
 \node at ({(\xLz+\xLu)/2},-3) {$u$};
 \node at ({(\xLu+\xLuv)/2},-3) {$v$};
}

{\tikzstyle{every path}=[draw,thick,-]
\path (\xLz,-1.0) -- (\xLu,-1.0) -- (\xLu,1.0) -- (\xLz,1.0) -- cycle;
\path (\xLu,-1.0) -- (\xLuv,-1.0) -- (\xLuv,1.0) -- (\xLu,1.0) -- cycle;
}
{\tikzstyle{every path}=[draw,dashed,-]
\path (\xLz,-1.3)   -- (\xLz,1);
\path (\xLu,-1.3)   -- (\xLu,1);
\path (\xLuv,-1.3)   -- (\xLuv,1);
\path (\xuo,-1.3)   -- (\xuo,1);
\path (\xui,-1.3)   -- (\xui,1);
\path (\xuii,-1.3)   -- (\xuii,1);
\path (\xumi,-1.3)   -- (\xumi,1);
\path (\xvmo,-1.3)   -- (\xvmo,1);
\path (\xvmi,-1.3)   -- (\xvmi,1);
\path (\xvi,-1.3)   -- (\xvi,1);
}
{\tikzstyle{every node}=[node font=\tiny]
\node at (\xLz,-2) {$0$};
\node at (\xLu,-2) {$|u|$};
\node at (\xLuv,-2) {$L$};
\node at (\xuo,-2) {$\alpha(0)$};
\node at (\xui,-2) {$\alpha^2(0)$};
\node at (\xuii,-2) {$\alpha^3(0)$};
\node at (\xumi,-2) {$\alpha^n(0)$};
\node at ({(\xuii+\xumi)/2},-2) {$\cdots$};
\node at (\xvmo,-2) {$\beta(L)$};
\node at (\xvmi,-2) {$\beta^2(L)$};
\node at (\xvi,-2) {$\beta^{n'}\!\!(L)$};
\node at ({(\xvi+\xvmi)/2},-2) {$\cdots$};
}

{\tikzstyle{every path}=[-Latex,shorten <=1mm]
\path (\xLz,1.2) edge [bend left=30] node [above] {$\alpha$} (\xuo,1.2);
\path (\xuo,1.2) edge [bend left=30] (\xui,1.2);
\path (\xui,1.2) edge [bend left=30] (\xuii,1.2);
\path (\xuii + 1.5,1.5) edge [bend left=30] (\xumi,1.2);
\path (\xLuv,1.2) edge [bend right=30] node [above] {$\beta$} (\xvmo,1.2);
\path (\xvmo,1.2) edge [bend right=30] (\xvmi,1.2);
\path (\xvmi,1.2) edge [bend right=30] (\xvmii,1.5);
\path (\xvi + 1.5,1.5) edge [bend right=30] (\xvi,1.2);
}
{\tikzstyle{every path}=[-Latex,shorten <=1mm,dash pattern=on \pgflinewidth off 1pt,bend left=40]
 \node[minimum size=0pt,font=\tiny] (tip1) at (\xLu + 0.85, 1.4){\textbf{?}};
 \node[minimum size=0pt,font=\tiny] (tip2) at (\xLu - 0.85, 1.4){\textbf{?}};
 \path (\xumi,1.2) edge [bend left=40] (\xLu + 0.7, 1.5);
 \path (\xvi,1.2) edge [bend right=40] (\xLu - 0.7, 1.5);
}

\end{tikzpicture}
 \end{center}
\caption{Comparing $\iota(u\, v)$ with $\iota(u)+\iota(v)$.}
\label{fig-alpha-for-uv}
\end{figure}
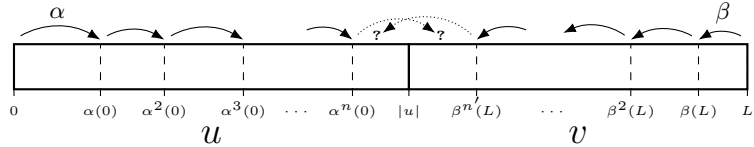

We can generalise \Cref{eq-iota-via-alpha}: $\iota(w)=n$ implies
$\alpha^p(0)\leq \beta^{n-p}(|w|)$ for all $p=0,\ldots,n$, and the
reciprocal holds.  We can use this to prove the following:
\begin{proposition}
$\iota(u \, v)\leq \iota(u)+\iota(v)+1$.
\end{proposition}
\begin{proof}
Write $n$ and $n'$ for $\iota(u)$ and $\iota(v)$.  Thus, on $w=u \, v$ with
$L=|u|+|v|$, one has $\alpha^{n+1}(0)>|u|$ and
$\beta^{n'+1}(L)<|u|$. See Fig.~\ref{fig-alpha-for-uv}. Hence
$\iota(w)<n+n'+2$.
\end{proof}
We can also reprove a result from~\cite{barker2020}:
\begin{proposition}
$\iota(u \, u^\Rrm)=2\iota(u)$.
\end{proposition}
\begin{proof}
Write $n$ for $\iota(u)$. When $w=u\,u^\Rrm$ and $L=|w|$, the factor
$w\bigl(\alpha^n(0),\beta^n(L)\bigr)$ is $r(u)\cdot r(u)^\Rrm$ hence
is not rich. Thus $\alpha^{n+1}(0)>|u|+|r(u)^\Rrm|=\beta^n(L)$, entailing
$\iota(u \, u^\Rrm)<2n+1$.
\end{proof}

\subsection{Subword circular universality index via jumping functions}

\begin{figure}[htbp]
\begin{center}
\scalebox{0.91}{\begin{tikzpicture}[auto,x=5mm,y=3mm,anchor=mid,baseline,node distance=1.3em]

{\tikzstyle{every node}=[]
\node at (1,0) {$s_1$};
\node at (3,0) {$s_2$};
\node at (5,0) {$s_3$};
\node at (7.5,0) {$\cdots$};
\node at (10,0) {$s_m$};
\node at (12,0) {$r$};
\node at (14,0) {$s_1$};
\node at (16,0) {$s_2$};
\node at (18,0) {$s_3$};
\node at (20.5,0) {$\cdots$};
\node at (23,0) {$s_m$};
\node at (25,0) {$r$};
}

{\tikzstyle{every path}=[draw,thick,-]
\path (0,-1.0) -- (13,-1.0) -- (13,1.0) -- (0,1.0) -- cycle;
\path[dashed] (2,-1.5) -- (2,1);
\path[dashed] (4,-1.5) -- (4,1);
\path[dashed] (6,-1) -- (6,1);
\path[dashed] (9,-1.5) -- (9,1);
\path[dashed] (11,-1.5) -- (11,1);
\path (13,-1.0) -- (26,-1.0) -- (26,1.0) -- (13,1.0) -- cycle;
\path[dashed] (15,-1.5) -- (15,1);
\path[dashed] (17,-1.5) -- (17,1);
\path[dashed] (19,-1) -- (19,1);
\path[dashed] (22,-1) -- (22,1);
\path[dashed] (24,-1) -- (24,1);

{\tikzstyle{every node}=[node font=\tiny]
\path (0,0) -- (0,-1.5);
\path (13,0) -- (13,-1.5);
\path (26,0) -- (26,-1.5);
\node at (0,-2) {$0$};
\node at (2,-2) {$\lambda_1$};
\node at (4,-2) {$\lambda_2$};
\node at (9,-2) {$\lambda_{m-1}$};
\node at (11,-2) {$\lambda_m$};
\node at (13,-2) {$\ell$};
\node at (15,-2) {$\lambda_1+\ell$};
\node at (17,-2) {$\lambda_2+\ell$};
\node at (26,-2) {$2\ell$};

\path (2.8,0.7) -- (2.8,1.15);
\path (15.8,0.7) -- (15.8,1.15);
\node at (2.8,1.5) {$i$};
\node at (15.8,1.5) {$i+\ell$};
}
}
{\tikzstyle{every node}=[node font=\Large]
\node at (6.5,-3) {$u$};
\node at (19.5,-3) {$u$};
}
{\tikzstyle{every path}=[thin,-Latex,shorten <=1mm]
\path (2.8,2) edge [bend left=30] node [above] {$\alpha$} (4.7,1.2);
\path (4.7,1.2) edge [bend left=30] (6.5,1.2);
\path (6.5,1.2) edge [dashed,bend left=30] (7.9,1.2);
\path (7.9,1.2) edge [dashed,bend left=30] (9.9,1.2);
\path (9.9,1.2) edge [dashed,bend left=30] (13.3,1.2);
\path (13.3,1.2) edge [bend left=30] (15.1,1.2);
}

\end{tikzpicture}
 }
\end{center}
\caption{Computing $\iota(u^{\sim i})$ on $w = u^2$.}
\label{fig-alpha-for-zeta}
\end{figure}
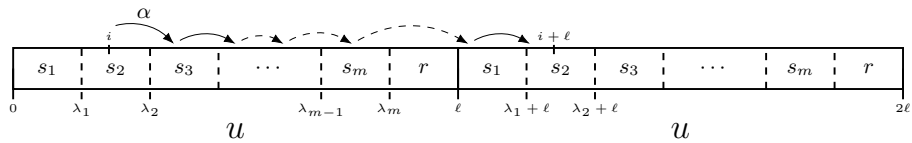

The jumping functions can be used to study the circular universality
index $\zeta(u)$. For this we consider the word $w = u\, u$ obtained
by concatenating two copies of $u$, so that $L = 2\ell$. Now, instead
of considering the conjugates of $u$, we can consider the factors
$w(i,i+\ell)$ of $w$:  see \Cref{fig-alpha-for-zeta}.

This leads to a characterisation of $\zeta(u)$ in terms of $\alpha$ on
$w = u \, u$:
\begin{align}
\label{eq-zeta-via-alpha-v0}
 \zeta(u)
         &= \max_{0\leq i<\ell} \: \max \bigl\{ n ~\big|~ \alpha^n(i)
 \leq i+\ell\bigr\}
\\
\shortintertext{or, using  $u^{\sim \ell}=u^{\sim 0}$,}
\label{eq-zeta-via-alpha}
         &= \max_{0< i\leq\ell} \: \max \bigl\{ n ~\big|~ \alpha^n(i) \leq i+\ell\bigr\}\:.
\end{align}

\paragraph{Bounding $\zeta(u)$.}
For $k=0,\ldots,m$, we write $\lambda_k$ for the cumulative length
$|s_1\cdots s_k|$ of the $k$ first arches of $u$, i.e., we let
$\lambda_k\eqdef\alpha^k(0)$.

The following Lemma and its corollary are a version of Lemma~20 from
\cite{barker2020} but we give a different proof.
\begin{lemma}
\label{lem-iota-conjugates}
Let $u$ and $u'$ be two conjugate words.
\\
(a) $\iota(u)-1\leq \iota(u')\leq\iota(u)+1$.
\\
(b) If furthermore $r(u)=\epsilon$ then $\iota(u')\leq \iota(u)$.
\end{lemma}
\begin{proof}
Let $s_1\cdots s_m\cdot r$ be the arch factorization of $u$ and assume that
$u'=u^{\sim i}$ as depicted in \Cref{fig-alpha-for-zeta}.
\\
(a) If the position $i$ falls inside some arch $s_p$ of $u$ (or inside the rest
$r$) we see that $s_{p+1}\cdots s_m\cdot s_1 \cdots s_{p-1}$ is a subword of
$u'$ hence $\iota(u') \geq m-1$.  This gives $\iota(u)-1 \leq \iota(u')$, and
the other inequality is obtained by exchanging the roles of $u$ and $u'$.  \\
(b) If furthermore $r=\epsilon$, then $\lambda_{p-1}\leq i<\lambda_p$ for some
$p$. Looking at $u'$ as a factor of $w=u^2$ (and assuming that
$\alpha^{m+1}(i)$ is defined) we deduce $\alpha^{m+1}(i) \geq
\alpha^{m+1}(\lambda_{p-1})=\lambda_p+\ell > i+\ell$. This proves
$\iota(u^{\sim i}) < m+1$.
\end{proof}
\begin{corollary}
\label{coro-zeta-iota}
(a) $\iota(u)\leq \zeta(u)\leq \iota(u)+1$.
\\
(b) Furthermore, if $r(u)=\epsilon$, then $\zeta(u)=\iota(u)$.
\end{corollary}

 \section{An $O(|u|\cdot|A|)$ algorithm for $\zeta(u)$}
\label{sec-zeta-algo}

The following crucial lemma shows that computing $\zeta(u)$ does not require
checking all the conjugates $u^{\sim i}$ for $0\leq i<\ell$.
\begin{lemma}
\label{lem-lambda1}
Let $u=a_1\cdots a_\ell$ be a rich word with arch factorization $s_1\cdots
s_m\cdot r$.
\\
(a) There exists some $0<d\leq \lambda_1 \eqdef |s_1|$ such that $\zeta(u) =
\iota(u^{\sim d})$.  \\
(b) Furthermore, there exists $a\in A$ such that $d = \min\{i~|~a_i=a\}$, i.e.,
$d$ can be chosen as a position right after a first occurrence of a letter in $u$.
\end{lemma}
\begin{proof}
Let $n = \zeta(u)$. For (a) it is enough to show that $\iota(u^{\sim
d})\geq n$ for some $d\in (0,\lambda_1]$.

By \Cref{eq-zeta-via-alpha} there exists some $0<i_0\leq\ell$ such that
$\alpha^n(i_0)\leq i_0+\ell$.  We consider the sequence $i_0 < i_1 < \cdots <
i_n$ given by $i_{k+1} = \alpha(i_k)$.  If $i_n \leq \ell$ then taking $d = 1$
works: monotonicity of $\alpha$ entails $\alpha^n(d) \leq \alpha^n(i_0) \leq
\ell$ and we deduce $\iota(u^{\sim d}) \geq n$. Clearly $d=1$ fulfils (b).

So assume $i_n>\ell$ and let $k$ be the largest index such that $i_k\leq \ell$
(hence $k<n$).  Since $\alpha(\ell) = \ell+\lambda_1$ (recall $\lambda_1 \eqdef
|s_1|$), monotonicity of $\alpha$ entails $i_{k+1} = \alpha(i_k) \leq
\ell+\lambda_1$, i.e., $i_{k+1}$ lands inside the first arch of the second copy
of $u$ in $w$.

Let now $d\eqdef i_{k+1}-\ell$ so that $u^{\sim d} = w(d,d+\ell) =
w(d,i_{k+1})$. Since $\alpha^{n-k-1}(i_{k+1}) = i_n\leq i_0+\ell$, one has
$\alpha^{n-k-1}(d) \leq i_0$ hence $\iota\bigl(w(d,i_0)\bigr) \geq n-k-1$.  We
also have $\iota\bigl(w(i_0,i_{k+1})\bigr) = k+1$ since $i_{k+1} =
\alpha^{k+1}(i_0)$.  This yields
\[
          \iota(u^{\sim d})
        = \iota\bigl(w(d,d+\ell)\bigr)
        \geq (n-k-1)  + (k+1)
        = n                         \:,
\]
entailing (a). For (b) observe that $w(i_{k+1}-1,i_{k+1})$ is the last letter
of an arch across the end of the first $u$ in $w$ to the beginning of the
second $u$ in $w$. Since it is the first occurrence of this letter in this
arch, it is also in $u$. Since $d$ is $i_{k+1}$ shifted to the first copy of
$u$, (b) is fulfilled.
\end{proof}

\begin{algorithm}[Computing $\zeta(u)$]
\label{algo-zeta-algo}
For each position $d$ such that $u(d-1,d)$ is the first occurrence of
a letter in $u$, one computes $\iota(u^{\sim d})$ (in time $O(|u|)$
for each $d$), and returns the maximum value found.
\qed
\end{algorithm}
The correctness of this algorithm is given by \Cref{lem-lambda1} (if
$u$ is not rich, $\zeta(u)=0$ and this will be found out during the
computation of $\iota(u^{\sim 1})$).  It runs in time $O(|A|\cdot
|u|)$ since there are at most $|A|$ values for $d$, starting with $d=1$.

There are two heuristic improvements that can speed up the
algorithm\footnote{They do not
improve the \emph{worst-case} complexity.}:
\begin{itemize}
\item As soon as we have encountered two different values
  $\iota(u^{\sim d}) \neq \iota(u^{\sim d'})$, we can stop the search
  for a maximum in view of \cref{coro-zeta-iota}.(a).

  For example, for $u=\ta\ta\tb\ta\tc\tc\tb$, the first occurrences of
  $\ta$, $\tb$, and $\tc$, are with $d=1$, $3$ and $5$. So one starts
  with computing $\iota(u^{\sim 1})=\iota(\ta\tb\ta\tc\tc\tb \,
  \ta)=2$. Then one computes $\iota(u^{\sim 3})=\iota(\ta\tc\tc\tb \,
  \ta\ta\tb)=1$. Now, and since we have encountered two different
  values, we may conclude immediately that $\zeta(u)=2$ without the
  need to compute $\iota(u^{\sim 5})$.

\item When computing some $\iota(u^{\sim d})$ leads us to notice
  $r(u^{\sim d}) = \epsilon$, we can stop the search in view of
  \cref{coro-zeta-iota}.(b).

  For example, and again with $u=\ta\ta\tb\ta\tc\tc\tb$, the
  computation of $\iota(u^{\sim 1})$ led us to the arch-factorization
  $u^{\sim 1} = \ta\tb\ta\tc \cdot \tc\tb\ta \cdot \epsilon$, with 2
  arches and with $r(u^{\sim 1})=\epsilon$. We may conclude
  immediately that $\zeta(u)=\iota(u^{\sim 1})=2$ without trying the
  remaining conjugates.
\end{itemize}

Observe that the above algorithm does not have to explicitly build $u^{\sim
d}$. It is easy to adapt any naive algorithm for $\iota(u)$ so that it starts
at some position $d$ and wraps around when reaching the end of $u$.

 \section{Subword universality signatures}
\label{sec-signatures}

In this section, we write $\iota_*(u)$, $r_*(u)$, etc., to denote the
values of $\iota(u)$, $r(u)$, etc., \emph{when one assumes that $A(u)$
is the underlying alphabet}. This notation is less heavy than writing,
e.g., $\iota_{A(u)}(u)$, but it is needed since we shall consider
simultaneously $\iota_*(u)$ and $\iota_*(v)$ when $A(u)\neq A(v)$,
i.e., when the two universality indexes have been obtained in
different contexts.

When $u$ is a word, we define a function $S_u$ on words via:
\begin{equation}
\label{eq-def-Su}
        S_u(x) = \bpair{\iota_*(x \, u), A\bigl(r_*(x \, u)\bigr)}
          \;\;\;\text{for all $x$ such that $A(u)\not\subseteq A(x)$.}
\end{equation}
In other words, $S_u(x)$ is a summary of the arch factorization of
$x\,u$: it records the number of arches in $x\,u$ and the letters of
the rest $r_*(x\,u)$, assuming that the alphabet is $A(x\,u)$.

Note that $S_u(x)$ is only defined when $A(u)\not\subseteq A(x)$,
i.e., when at least one letter from $u$ does not appear in $x$. With
this restriction, $S_u(x)$ and $S_u(x')$ coincide (or are both
undefined) whenever $A(x)=A(x')$.  For this reason, we sometimes write
$S_u(B)$, where $B$ is a set of letters, to denote any $S_u(x)$ with
$A(x)=B$.

\medskip

We are now almost ready to introduce the main new object: a compact
data structure with enough information for computing $S_u$ on
arbitrary arguments.

With a word $u$ we associate $e(u)$, a word listing the letters of $u$ in the
order of their first appearance in $u$. For example, by underlining the first
occurrence of each letter in $u = \mathtt{\underline{c} c \underline{a} c a
\underline{b} c b b a}$ we show $e(u) = \mathtt{c a b}$. We also write $f(u)$
for the word listing the letters of $u$ in order of their last
occurrence: in the previous example $f(u)= \mathtt{c b a}$.

\begin{definition}
The \emph{subword universality signature} of a word $u$ is the pair
$\Sigma(u)=\pair{e(u),\tS_u}$ where  $\tS_u$ is
$S_u$ restricted to the strict suffixes of $e(u)$.
\end{definition}
\begin{example}
With $u=\mathtt{\underline{a} a \underline{b} a \underline{c}}$ we have:
\[
\Sigma(u) =
\left\{
\begin{array}{l}
e(u) = \ta\tb\tc \\[.4em]
\tS_u =
\left\{\begin{array}{rl}
 \epsilon &\mapsto \bpair{1, \emptyset   }  \\
 \tc      &\mapsto \bpair{1, \{\ta,\tc\} }  \\
 \tb\tc   &\mapsto \bpair{2, \emptyset   }
\end{array}\right.\;\;\; \text{ in view of: } \; \;
\begin{array}{rl}
\epsilon \cdot u &= \mathtt{a a b a c} \cdot \epsilon                \\
     \tc \cdot u &= \mathtt{c a a b}   \cdot \mathtt{a c}            \\
 \tb \tc \cdot u &= \mathtt{b c a}   \cdot \mathtt{a b a c} \cdot \epsilon
\end{array}
\end{array}
\right.
\]
NB: the strict suffixes of $e(u)$ are $\epsilon$, $\tc$
and $\tb\tc$.
\end{example}

While finite (and quite small) $\Sigma(u)$ contains enough information
for computing $S_u$ on any argument $x$ on any alphabet. One can use
the following algorithm:
\begin{algorithm}[Computing $S_u(x)$ from $\Sigma(u)$]
\label{algo-Su}~\\
Given inputs $x$ and $\Sigma(u)=\pair{e(u),\tS_u}$ we proceed as follows:
\\
(a) Retrieve $A(u)$ from $e(u)$. Check that $A(u)\not\subseteq
A(x)$, since otherwise $S_u(x)$ is undefined.\\
(b) Now with $x\in\dom(S_u)$, let $y$ be the longest suffix of $e(u)$
with $A(y)\subseteq A(x)$ ---necessarily $y$ is a \emph{strict}
suffix of $e(u)$--- and extract $\pair{n_y,B_y}$ from $\tS_u(y)$.\\
(c.1) If $A(x)\subseteq A(u)$, return $S_u(x)=\pair{n_y,B_y}$.
\\
(c.2) Similarly, if $n_y=1$ return $S_u(x)=\pair{n_y,B_y}$.
\\
(c.3) Otherwise return $S_u(x)=\pair{1,A(u)}$.
\end{algorithm}
\begin{proof}[of correctness]
Assume $x\in\dom(S_u)$.  Since $u$ contains a letter not appearing in
$x$, the first arch of $x\, u$ ends inside $u$, so let us consider the
factorization $u=u_1u_2$ such that $x\, u_1$ is the first arch of $x
\, u$ (see picture below, where $e(u)$ is underlined).
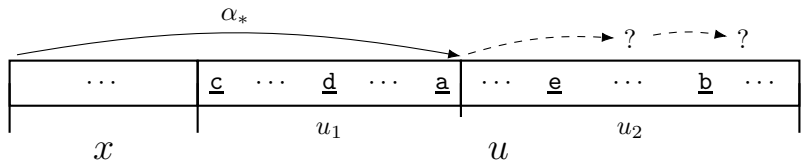
\begin{center}
\begin{tikzpicture}[auto,x=5mm,y=3mm,anchor=mid,baseline,node distance=1.3em]

{\tikzstyle{every node}=[]
\node at (2.5,0) {$\cdots$};
\node at (5.5,0) {$\underline{\mathtt{c}}$};
\node at (7.0,0) {$\cdots$};
\node at (8.5,0) {$\underline{\mathtt{d}}$};
\node at (10.0,0) {$\cdots$};
\node at (11.5,0) {$\underline{\mathtt{a}}$};
\node at (13.0,0) {$\cdots$};
\node at (14.5,0) {$\underline{\mathtt{e}}$};
\node at (16.5,0) {$\cdots$};
\node at (18.5,0) {$\underline{\mathtt{b}}$};
\node at (20.0,0) {$\cdots$};
}

{\tikzstyle{every path}=[draw,thick,-]
\path (0,-1.0) -- (5,-1.0) -- (5,1.0) -- (0,1.0) -- cycle;     \path (5,-1.0) -- (12,-1.0) -- (12,1.0) -- (5,1.0) -- cycle;   \path (12,-1.0) -- (21,-1.0) -- (21,1.0) -- (12,1.0) -- cycle; 

{\tikzstyle{every node}=[node font=\tiny]
\path (0,0) -- (0,-2.2);
\path (5,0) -- (5,-2.2);
\path (12,0) -- (12,-1.5);
\path (21,0) -- (21,-2.2);
}
}
{\tikzstyle{every node}=[node font=\Large]
\node at (2.5,-3) {$x$};
\node at (13,-3) {$u$};
}
{\tikzstyle{every node}=[]
\node at (8.5,-2) {$u_1$};
\node at (16.5,-2) {$u_2$};
}
{\tikzstyle{every path}=[thin,-Latex,shorten <=1mm]
\path (0,1.2) edge [bend left=10] node [above] {$\alpha_*$} (12,1.2);
\node (qm) at (16.5,1.8) {?};
\path (12,1.2) edge [dashed,bend left=10] (qm);
\node (qm2) at (19.5,1.8) {?};
\path (qm) edge [dashed,bend left=10] (qm2);
}

\end{tikzpicture}
 \end{center}
Now $u_1$ has a last letter, say $a$, that appears only once in $u_1$
and not at all in $x$.  Observe that a letter $b$ appears after $a$ in
$e(u)$ iff it does not appear in $u_1$, and thus must appear in
$x$. Hence the $y$ computed in step (b) is the suffix of $e(u)$ after
$a$ (in the above picture $y$ would be $\mathtt{eb}$).

If $A(x)\subseteq A(u)$ then $y \, u_1$ is rich, and is in fact an
arch since its last letter, $a$, appears only once. So $S_u(x)$ and
$S_u(y)$ coincide and step (c.1) is correct.

In case $A(x)\not\subseteq A(u)$, both $x$ and $u$ contain some
letters that are absent from the other word, so necessarily $\iota_*(x
\, u)=1$ and $r_*(x\, u)=u_2$. There only remains to compute $A(u_2)$
from $\Sigma(u)$.  We know that $\tS_u(y)=\pair{n_y,B_y}$. If $n_y>1$
this means that $u_2$ contains at least another $A(u)$-arch, so
$A(u_2)=A(u)$ and step (c.3) is correct.  If $n_y=1$ this means that
$y\, u$ only has one arch, namely $y\, u_1$, and $B_y$ provides
$A(u_2)$: step (c.2) is correct in this case.
\end{proof}
\begin{remark}[Space and time complexity for \Cref{algo-Su}]
For simplifying our complexity evaluation, we assume that there is a
fixed maximum size for alphabets so that storing a letter $a\in A$
uses space $O(1)$, e.g., 64 bits.  When storing $\Sigma(u)$, the $e(u)$
part uses space $O(|A|)$. Now $\tS_u$ can be represented in space
$O(|A|\log |u|)$ when $e(u)$ and $f(u)$ are known: it contains at most
$|A|$ pairs $\pair{n_x,B_x}$ where $x$ is a suffix of $e(u)$ and $B_x$
is always the alphabet of a strict suffix of $f(u)$: $x$ and $B_x$ can
thus be represented by a position (or a letter) in $e(u)$ and
$f(u)$. The $n_x$ values each need at most $\log |u|$ bits.

Regarding time, the algorithm runs in time
$O\bigl(|x|+|\Sigma(u)|+|A(u)|\bigr)$.
\qed
\end{remark}

\subsection{Universality indexes from signatures}

Obviously the signature $\Sigma(u)$ contains enough information for retrieving
$\iota_*(u)$: this is found in $\tS_u(\epsilon)$.  More interestingly, one can
also retrieve $\zeta_*(u)$:
\begin{proposition}
\label{prop-zeta-from-sig}
Let $u$ be a word with $\iota_*(u) = m$. Then $\zeta_*(u) = m+1$ iff there
exists a strict suffix $x$ of $e(u)$ with $\tS_u(x) = \pair{n_x,B_x}$ such that
$n_x = m+1$ and $A(x) \subseteq B_x$. Otherwise $\zeta_*(u) = m$.
\end{proposition}
\begin{proof}
($\Leftarrow$): assume $\tS_u(x)=\pair{m+1,B_x}$ with $A(x)\subseteq B_x$.
Thus $\iota_*(x\, u)=m+1$. Factor $u$ as $u=u_1u_2r$ such that $x\, u_1$ is the
first arch of $x\, u$ and such that $r=r_*(x\, u)$ is its rest.  Then $u_2$
contains $m$ arches and $B_x=A(r)$.  Let now $u'\eqdef r\, u_1u_2$.  We claim
that $\iota_*(u')= m+1$. Indeed $r\, u_1$ is rich since $x\, u_1$ is rich and
$A(x) \subseteq A(r)$, so $\iota_*(r\,u_1u_2) \geq m+1$. Since $u'$ and $u$ are
conjugates, we deduce $\zeta_*(u) = \iota_*(u') = m+1$ from
\Cref{coro-zeta-iota}.(a).\\
($\Rightarrow$): assume $\zeta_*(u) = m+1$. By \Cref{lem-lambda1} we know that
$\iota_*\bigl(u^{\sim i}\bigr) = m+1$ for some position $0 < i \leq \lambda_1$
falling just after a first occurrence of a letter in $u$. Looking at factors of
$w = u\, u$ as we did before, we have $\alpha^{m+1}(i) \leq i+\ell$, leading to
$j \eqdef \alpha^m(i) \leq \ell$ (see picture below).\\
\scalebox{0.91}
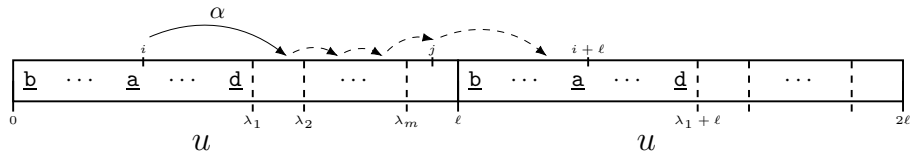
{\begin{tikzpicture}[auto,x=5mm,y=3mm,anchor=mid,baseline,node distance=1.3em]

{\tikzstyle{every node}=[]
\node at (0.5,0) {$\underline{\mathtt{b}}$};
\node at (2,0) {$\cdots$};
\node at (3.5,0) {$\underline{\mathtt{a}}$};
\node at (5,0) {$\cdots$};
\node at (6.5,0) {$\underline{\mathtt{d}}$};
\node at (10,0) {$\cdots$};
\node at (13.5,0) {$\underline{\mathtt{b}}$};
\node at (15,0) {$\cdots$};
\node at (16.5,0) {$\underline{\mathtt{a}}$};
\node at (18,0) {$\cdots$};
\node at (19.5,0) {$\underline{\mathtt{d}}$};
\node at (23,0) {$\cdots$};
}

{\tikzstyle{every path}=[draw,thick,-]
\path (0,-1.0) -- (13,-1.0) -- (13,1.0) -- (0,1.0) -- cycle;
\path[dashed] (7,-1.5) -- (7,1);\path[dashed] (8.5,-1.5) -- (8.5,1);\path[dashed] (11.5,-1.5) -- (11.5,1);\path (13,-1.0) -- (26,-1.0) -- (26,1.0) -- (13,1.0) -- cycle;
\path[dashed] (20,-1.5) -- (20,1);\path[dashed] (21.5,-1.5) -- (21.5,1);\path[dashed] (24.5,-1.5) -- (24.5,1);

{\tikzstyle{every node}=[node font=\tiny]
\path (0,0) -- (0,-1.5);
\path (13,0) -- (13,-1.5);
\path (26,0) -- (26,-1.5);
\node at (0,-2) {$0$};
\node at (7,-2) {$\lambda_1$};\node at (8.5,-2) {$\lambda_2$};
\node at (11.5,-2) {$\lambda_m$};
\node at (13,-2) {$\ell$};
\node at (20,-2) {$\lambda_1+\ell$};
\node at (26,-2) {$2\ell$};

\path (3.8,0.7) -- (3.8,1.15);\node at (3.8,1.5) {$i$};
\path (12.25,0.7) -- (12.25,1.15);\node at (12.25,1.5) {$j$};
\path (16.8,0.7) -- (16.8,1.15);
\node at (16.8,1.5) {$i+\ell$};
}
}
{\tikzstyle{every node}=[node font=\Large]
\node at (5.5,-3) {$u$};
\node at (18.5,-3) {$u$};
}
{\tikzstyle{every path}=[thin,-Latex,shorten <=1mm]
\path (3.8,2) edge [bend left=30] node [above] {$\alpha$} (8,1.2);
\path (8,1.2) edge [dashed,bend left=30] (9.5,1.2);
\path (9.5,1.2) edge [dashed,bend left=30] (10.8,1.2);
\path (10.8,1.2) edge [dashed,bend left=30] (12.3,2);
\path (12.3,2) edge [dashed,bend left=30] (15.6,1.2);
}

\end{tikzpicture}
 }\\
Define now $x$ as the suffix of $e(u)$ that contains all letters in
$u(i,\lambda_1)$, that is, all underlined letters to the right of $i$. This is
a strict suffix since $i>0$. Now $x\,u(0,i)$ is rich, and $u(i,j)$ is made of
exactly $m$ arches, so $\iota_*(x\,u) = n_x = m+1$ and $r_*(x\, u)=u(j,\ell)$.
\\
Then $B_x = A\bigl(u(j,\ell)\bigr)$ and $w(j,i+\ell)$ is rich, so
$w(j,\ell)$  contains all letters missing from $w(i,i+\ell)=u(0,i)$. In
other words $B_x\supseteq A(x)$,  concluding the proof.
\end{proof}
\begin{corollary}[Computing universality indexes from signatures]
\label{algo-zeta-from-sig}
One can compute $\iota_*(u)$ and $\zeta_*(u)$ from $\Sigma(u)$ in time
$(|A|+\log|u|)^{O(1)}$.
\end{corollary}
Actual implementations can use heuristics based on
\Cref{lem-iota-conjugates}.(b): if $\tS_u(\epsilon)=\pair{m,\emptyset}$ then
$\zeta_*(u)=m$.

 \subsection{Combining signatures}
\label{sec-combi-sig}

Subword universality signatures can be computed compositionally.
\begin{algorithm}[Combining signatures]
\label{algo-comp-sig}
The following algorithm takes as input the signatures $\Sigma(u)$ and
$\Sigma(v)$ of any two words and computes $\Sigma(u\, v)$:
\\

\noindent
(a) Retrieve $A(u)$ and $A(v)$ from $e(u)$ and $e(v)$, then compute
$e(u\,v)$ as $e(u)\,e'$ where $e'$ is the subword of $e(v)$ that only
retains the letters from $A(v)\setminus A(u)$.
\\
(b) Consider now any strict suffix $x$ of $e(u \, v)$ and compute $\tS_{u\,
v}(x)$ as follows:
\\
(b.1) If $A(v)\not\subseteq A(x)\cup A(u)$ then let $\tS_{u\, v}(x)\eqdef
S_v\bigl(x\,e(u)\bigr)$,  using \Cref{algo-Su}.
\\
(b.2) If $A(v)\subseteq A(x)\cup A(u)$, then $A(u)\not\subseteq A(x)$.
Write $\bpair{n,B}$ for $\tS_u(x)$:
\\
(b.2.1) If now $A(v)\cup B\neq A(x)\cup A(u)$ then let $\tS_{u\, v}(x)\eqdef
\bpair{n,A(v)\cup B}$.
\\
(b.2.2) Otherwise retrieve $\tS_v(B) = \bpair{n',B'}$ and let $\tS_{u\,
v}(x)\eqdef \bpair{n+n',B'}$.
\end{algorithm}
\begin{proof}[of correctness]
Step (a) for $e(u\, v)$ is correct.
\\
In step (b) we want to compute $S_{u\, v}(x)$. Now $x\,(u\,v)=(x\,
u)\,v$ so $S_{u\, v}(x)$ coincides with $S_v(x \, u)$ \emph{when the
latter is defined} . This is the case in step (b.1) where one computes
$S_v(x\,u)$ by replacing $x\, u$ with $x\, e(u)$, an argument with
same alphabet (recall that the algorithm does not have access to $u$
itself).  \\
In step (b.2) where $S_v(x\,u)$ is not defined, computing $S_u(x)$ provides
$n$ and $B=A(r)$ for the arch factorization $x\,u=s_1\cdots s_n\cdot r$ of
$x\,u$.\\
We can continue with the arch factorization of $r\,v$ and combine the two sets
of arches if these factorizations rely on the same alphabet: this is step
(b.2.2).  \\
Otherwise, $r\,v$ only uses a subset of the letters of $x\, u$. There won't be
a new arch, only a longer rest: $r_*(x\,u\,v)=r\,v$. Step (b.2.1) is correct.
\end{proof}
Note that \Cref{algo-comp-sig} runs in time $O\bigl(|A(u\,
v)|+|\Sigma(u)|+|\Sigma(v)|\bigr)$ and that the result has linear
size $|\Sigma(u\, v)|=O(|\Sigma(u)|+|\Sigma(v)|)$.

\section{Universality indexes for SLP-compressed words}
\label{sec-SLPs}

We are now ready to compute the universality indexes of SLP-compressed words.
Recall that an SLP $X$ is an acyclic context-free grammar in Chomsky normal
form where furthermore each non-terminal has only one production rule, i.e.,
the grammar is deterministic (see survey~\cite{lohrey2012}).  SLPs are the
standard mathematical model for compression of texts and files and, modulo
polynomial-time encodings, it encompasses most compression schemes used in
practice.

Formally, an SLP $X$ with $m$ rules is a list $\pair{N_1\to \RHS_1;
\cdots; N_m\to \RHS_m}$ of production rules where each right-hand side
$\RHS_i$ is either a letter $a$ from $A$ or a concatenation
$N_{j}\,N_{j'}$ of two nonterminals with $j,j'<i$.  It has size
$|X|=O(m\log m)$ when $A$ is fixed.

Each nonterminal $N_i$ encodes a word, its \emph{expansion},
given inductively via:
\[
\exp(N_i)\eqdef
\begin{cases}
a
&\text{if $\RHS_i=a$,}
\\
\exp(N_j)\exp(N_{j'})
&\text{if $\RHS_i=N_j\, N_{j'}$.}
\end{cases}
\]
Finally, the expansion $\exp(X)$ of the SLP itself is the expansion
$\exp(N_m)$ of its last nonterminal. This is a word (or file) of length
$2^{O(|X|)}$ and one of the main goals in the area of compressed data
science is to develop efficient methods for computing relevant
information about $\exp(X)$ directly from $X$, i.e., without actually
decompressing the word or file.

In this spirit we can state:
\begin{theorem}
The universality indexes $\iota\bigl(\exp(X)\bigr)$ and
$\zeta\bigl(\exp(X)\bigr)$ can be computed from an SLP $X$ in
bilinear time $O\bigl(|A|\cdot|X|\bigr)$.
\end{theorem}
\begin{proof}
One just computes $\Sigma\bigl(\exp(N_1)\bigr)$, \ldots,
$\Sigma\bigl(\exp(N_k)\bigr)$ for the non-terminals $N_1,\ldots,N_k$ of
$X$. If $N_i$ is associated with a production rule $N_i\to
N_{i_1}N_{i_2}$, we compute $\Sigma\bigl(\exp(N_i)\bigr)$ by combining
$\Sigma\bigl(\exp(N_{i_1})\bigr)$ and $\Sigma\bigl(\exp(N_{i_2})\bigr)$ via
\Cref{algo-comp-sig} (recall that $i_1,i_2<i$ since the grammar is
acyclic). If $N_i$ is associated with a production $N_i\to a$ for some
$a\in A$, then $\Sigma\bigl(\exp(N_i)\bigr)=\Sigma(a)$ is trivial. In the
end we can extract the universality indexes of $\exp(X)$, defined as
$\exp(N_k)$, from $\Sigma\bigl(\exp(N_k)\bigr)$ using
\Cref{algo-zeta-from-sig}.  Note that all signatures have size
$O(|A|\cdot|X|)$ since for any $u=\exp(N_i)$, $\log |u|$ is in
$O(|X|)$. With the analysis of \Cref{algo-comp-sig} and
\Cref{algo-zeta-from-sig}, this justifies the claim about complexity.
\end{proof}

 \section{Conclusion}
\label{sec-conclusion}

We introduced arch-jumping functions and used them to describe and
analyse the subword universality and circular universality indexes
$\iota(u)$ and $\zeta(u)$. In particular,  this leads to a simple and
elegant algorithm for computing $\zeta(u)$.

In a second part we defined the subword universality signatures of
words, a compact data structure with enough information for
extracting $\iota(u)$ and $\zeta(u)$.  Since one can efficiently
compute the signature of $u\,v$ by composing the signatures of $u$ and
$v$, we obtain a polynomial-time algorithm for computing $\iota(X)$
and $\zeta(X)$ when $X$ is a SLP-compressed word. This raises our hopes
that one can compute some subword-based descriptive complexity
measures on compressed words, despite the known difficulties
encountered when reasoning about subwords.

\bibliographystyle{alpha}
\bibliography{../../BIBLIO/biblio}

\end{document}